\documentclass[a4paper,fleqn,usenatbib]{mnras}

\usepackage{mathtools}
\usepackage{bm}
\usepackage[T1]{fontenc}
\usepackage{ae,aecompl}
\usepackage{graphicx}

\def\apjs{ApJ. S}

\def\apjl{ApJL}

\def\aap{A\&A}

\newif\ifAMStwofonts
\AMStwofontstrue

\def\xid{{\cal D}}
\def\xis{{\cal S}}
\def\xidv{{\cal P}}

\def\vv{\bf v}

\def\vB{{\bf B}}

\def\gsim{~\rlap{$>$}{\lower 1.0ex\hbox{$\sim$}}}

\def\simpropto{\lower.2ex\hbox{$\; \buildrel \propto \over \sim \;$}}
\def\ltsim{\lower.5ex\hbox{$\; \buildrel < \over \sim \;$}}
\def\gtsim{\lower.5ex\hbox{$\; \buildrel > \over \sim \;$}}
\def\ltsim{\lower.5ex\hbox{$\; \buildrel < \over \sim \;$}}
\def\gtsim{\lower.5ex\hbox{$\; \buildrel > \over \sim \;$}}

\def\kms{\mbox{km\,s$^{-1}$}}

\def\dd{\,{\rm d}}

\def\kms{\ {\rm km\,s^{-1}}}
\def\hmpc{\ {\rm h^{-1}Mpc}}

\def\dd{{\rm d}}

\def\ln{{\rm ln}}

\def\la{\langle}
\def\ra{\rangle}
\def\grad{\nabla}
\def\pmb#1{\setbox0=\hbox{#1}%
\kern-.025em\copy0\kern-\wd0
\kern.05em\copy0\kern-\wd0
\kern-.025em\raise.0433em\box0}
\def\mb#1{\setbox0=\hbox{$#1$}%
\kern-.025em\copy0\kern-\wd0
\kern.05em\copy0\kern-\wd0
\kern-.025em\raise.0433em\box0}

\def\vv{\pmb{$v$}}

\def\vx{\pmb{$x$}}
\def\vX{\pmb{$X$}}
\def\vSigma{\pmb{$\Sigma$}}
\def\vxi{\pmb{$\xi$}}
\def\vDelta{\pmb{$\Delta$}}
\def\vr{\pmb{$r$}}
\def\vd{\pmb{$d$}}

\def\hvn{\hat {\vr}}
\def\hn{\hat {n}}
\def\hvx{\hat {\vx}}
\def\vk{\pmb{$k$}}

\def\simlt{\lower.5ex\hbox{$\; \buildrel < \over \sim \;$}}
\def\simgt{\lower.5ex\hbox{$\; \buildrel > \over \sim \;$}}

\newcommand{\beq}{\begin{equation}}
\newcommand{\eeq}{\end{equation}}
\def\beqa{\begin{eqnarray}}
\def\eeqa{\end{eqnarray}}
\def\fixit#1{}
\def\hmpc{h^{-1}\,{\rm Mpc}}

\def\dd{{\rm d}}

\def\apjs{ApJ. S}

\def\apjl{ApJL}

\def\aap{A\&A}

\usepackage{enumitem}
\title[ Bulk flow estimation]{On methods of  estimating cosmological  bulk flows}

  \author[A. Nusser]{
Adi Nusser,$^{1}$\thanks{E-mail: adi@physics.technion.ac.il}\\
$^{1}$Physics Department and the Asher Space Science Institute-Technion, Haifa 32000, Israel\\
}
\date{\today}

\begin{document}
    \label{firstpage}
\pagerange{\pageref{firstpage}--\pageref{lastpage}}
\maketitle

  \begin{abstract}

We explore  similarities and differences between  several  estimators of the cosmological bulk flow, $\vB$, from the observed radial peculiar velocities of galaxies.
A distinction is made between two  theoretical definitions of $\vB$ as a  dipole moment of the velocity field weighted by a radial window function. 
One definition involves the  three dimensional (3D) peculiar velocity, while the other is based on  its  radial component alone. 
Different methods attempt at inferring $\vB$ for either of these definitions  which coincide only for a constant velocity field. 
We focus on the  Wiener Filtering \citep[WF,][]{Hoffman15}  and  the Constrained Minimum Variance \citep[CMV,][]{feldwh10} methodologies. 
Both methodologies require a prior expressed in terms of the radial  velocity correlation function. 
\cite{Hoffman15} compute $\vB$ in Top-Hat windows from a  WF  realization of the 3D peculiar velocity field.
  \cite{feldwh10}  infer $\vB$ directly from 
the observed velocities  for the  second definition of $\vB$. 
The WF methodology could easily be 
adapted to the second definition, in which case it will be  equivalent to the CMV with the exception of the imposed constraint. 
For a  prior with vanishing correlations or very noisy data, CMV  reproduces the standard  Maximum Likelihood   \citep[ML,][]{Kaiser88}) estimation for $\vB$ of the entire sample independent of the  radial weighting function. 
Therefore, this estimator   is likely  more susceptible to observational  biases that could be present in measurements of distant galaxies. 
Finally, two additional estimators are proposed.

 \end{abstract}

\begin{keywords}
Cosmology: large-scale structure of the Universe, dark matter
\end{keywords}
%\pagebreak

\section{Introduction}
\label{sec:int}

On large scales where baryonic processes are dynamically unimportant, peculiar motions of galaxies are likely to be unbiased tracers of the 
peculiar velocity field of the dominant dark matter. This is in contrast to the spatial distribution of galaxies
 which is naturally biased with respect to the underlying mass density field. 
 Therefore,  catalogs of radial peculiar motions should offer a unique window to any dynamical deviations from  standard 
 gravity which may arise in theories for the observed cosmic acceleration \citep[e.g.][]{Hellwing2014}.  
One important caveat is that the  extraction of  cosmological information from  peculiar velocity catalogs generally involves 
the biasing relation between galaxies and  mass and also observational cuts imposed on the data, e.g. the Zone of Avoidance.   
 An example is a direct calculation of the  velocity correlation function where the desired signal 
 is modulated by how galaxies are distributed in the particular velocity catalog \citep[e.g.][]{Davis1983}. 
 A large scale moment of the  velocity field which has been the subject of  vivacious  debate  is the {\it bulk flow }. 
 Although  its estimation is affected by the spatial coverage of the data, the bulk flow is fairly insensitive to galaxy-mass biasing \cite{Li2012},  
 Let $\vv(\vr)$ be the three dimensional (3D) 
 peculiar velocity as a function of the comoving distance coordinate $\vr$ (in $ \kms$) and $u(\vr)=\vv \cdot \hvn$ be the corresponding radial 
 peculiar velocity field. We consider  two theoretical definitions of the bulk flow, $\vB$,
%\cite{Li2012}
\begin{equation}
\label{eq:Bdefone}
B^\alpha_I=\int \dd^3 r g(r) v^\alpha (\vr)  \; ,
\end{equation}
and 
\begin{equation}
\label{eq:Bdeftwo}
B_{II}^\alpha={3} \int \dd^3 r g(r) u(\vr) \hn^\alpha(\vr) \; ,
\end{equation} 
where $\hn^\alpha=\hvn\cdot \hvx^\alpha$ are cosine  angles between $\vr$ and  the  Cartesian axes defined by  unit vectors $\hvx^\alpha$ ($\alpha=1,2$ \& $3$).
The radial weighting function $g(r) $  is usually introduced in
the definition of $\vB$, and it satisfies  $\int \dd^3 g(r)=1$. 
For a Top-Hat window of radius $R$ centered on the observer $g=3/(4\pi R^3)$ for $r>R$ and vanishes otherwise. 
For a Gaussian window $g\propto \exp(-r^2/2R^2)$.  For  a constant 3D peculiar velocity, $\vv=\vB_0=const$, the two definitions yield the same result, i.e. $\vB_0$.
However, they do not coincide in general  \citep{Nusser2014a}. 

Any moment  of the peculiar  velocity field can serve as a basis for 
a quantity that can be estimated from the data for the purpose of constraining cosmological models. 
The general framework for the various usages of the term  bulk flow is the fact that the weighting function $g$ 
is independent of direction. 
 
 We aim at clarifying  the connection between methods for inferring  bulk  flows from sparse and noisy velocity 
 catalogs.  Different methods 
 adopt one  of the bulk flow definitions above \citep[e.g.][]{feldwh10,Hoffman15}. The lack of equivalence between the two definitions even for a full 3D velocity field, 
 guarantees  different results for some of the methods. 
 Further, although not always explicitly stated in the relevant papers, all methods rely on 
certain assumptions on how the observed radial motions are related to the full 3D velocity field. 
The bulk flow is defined as integrals over space including regions uncovered by the sparse data. Therefore, even in the ideal case of no observational errors, the data on its own is insufficient to compute $\vB$. Supplementing the missing information requires extrapolating the observed peculiar velocities to unobserved points in space.  
This is best done by   resorting to the  statistical nature of the  3D velocity field within the context of a cosmological model. A unwarranted criticism that is often raised is 
that the approach is  circular 
in the sense that the inferred $\vB$ 
 is necessarily consistent with the assumed model.  The target quantity (e.g. $\vB$) is certainly poorly constraint by very noisy data, in which case the prior information dominates.   
 Nevertheless, current catalogs, e.g.  the SFI++ \citep{Springob2007} and Cosmicflows-2 \citep[hereafter CF2,][]{CF2},  are sufficiently accurate and dense  and they constrain the bulk flow within $\sim 70\hmpc$ with very little dependence on the assumed prior  \citep{Nusser2011}.  
The methods discussed here will 
be addressed within the context 
of 
 gaussian random fields. 
The assumption is justified since  motions on large scales obey  linear theory for gravitational instability and hence, in the standard paradigm, the velocity and density fields remain  guassian.
We further have to specify the statistical nature of the 
random error on the estimated peculiar velocity.  Most common distance indicators
are  intrinsic scaling relations between galaxy observables involving the distance modulus (log distance) rather than the actual distance. 
Thus a normal (gaussian) scatter in these relations propagates into a skewed distribution of the measured distance (and peculiar velocity) error \citep[e.g.][]{Lynden-Bell1988,Springob2014}. 
The challenge of dealing with non-gaussian errors can be alleviated by
following the scheme of   \cite{Nusser1995,Nusser2011}.  Consider the (inverse)  Tully-Fisher relation as a distance indicator:
$\eta=s M-\eta+\eta_0+\Delta \eta$  where $s $ and $\eta_0$ are constants, $\eta$ is log the line-width, $M$ and $\Delta \eta $ is a random scatter with a gaussian distribution.  For $z\ll 1$, we have $cz=r+u$ and 
 $M=m-5\log(r) =M_0-5\log(1-u/cz)$, where $m$ is the apparent magnitude and $ M_0=m-5\log(cz)$ 
 Typically, $u/cz\ll 1$ even for nearby galaxies if $u$ is measured in Local Group frame. Nusser \& Davis adopt $ \Delta \eta= s M_0 +2.17s u/cz-\eta+\eta_0$ as the estimator for the velocity where $u$ is 
typically expressed in terms of a velocity model such as a bulk flow. This is equivalent to  setting
$u=0.46 cz  (\eta-\eta_0-s M_0)/s$ as the estimate of the peculiar velocity with a normally distributed random error.
A closely related  scheme has been advocated more  recently by \cite{Watkins2015} for obtaining  individual peculiar velocities.

We also ignore here any systematic biases in the determination of the peculiar velocities. To mitigate  spatial Malmquist biases \citep[e.g.][]{Lynden-Bell1988,Strauss1995a} we assume that galaxies are placed at their redshift coordinates rather than the observed distance \citep{Aaronson1982}. To linear order 
this is  equivalent to peculiar velocities expressed in terms of actual distance and hence, for simplicity of notation, we shall assume that the velocities are 
give in terms of the actual distance. 

The outline of the paper is as follows. The notation and the statistical tools are given in \S\ref{sec:pre}. The ``frequentist" approach to inferring $\vB$ is presented in \S\ref{sec:freq}. A new generalized ML estimation is presented here and the standard ML is shown to be recovered as a special case. 
In \S\ref{sec:bay} we describe the relevant  Baysian inference approach.  \S{sec:MV} focuses on  minimum variance and constrained minimum variance. This section includes  new estimator which  incorporates constraints in a probabilistic manner. A specific scheme for computing the relevant covariance matrices is given in \S\ref{sec:computing}. We end with a general discussion in \S\ref{sec:disc}.
%The paper relies on \cite{Kaiser88,Fisher95b,Zh95,feldwh10,Nusser2011,Agarwal2012,hr91,Hoffman15}
\section{Preliminaries}
\label{sec:pre}
\subsection{Notation}

We are provided with a  set of $N$ data points $d_i$ representing observations of the radial
peculiar velocities of galaxies, 
\begin{equation}
{d}_i=u_i+e_i\; ,
\end{equation}
where $u_i $ is the underlying true signal and $e_i$  is a random variable representing observational errors.
In general $\sigma_i^2=\sigma_*^2+\sigma_d^2$ where $\sigma_*$ correspond to small scale velocity dispersion and $\sigma_d $ 
is proportional to the distance and it arises from the intrinsic scatter in the distance indicator (e.g the Tully-Fisher relation). 
% The radial direction is designated by the unit vector $\hvn$ while  $\hvx^\alpha$ ($\alpha=1,2,3)$ are fixed unit vectors along the cartesian coordinates.
The cosine  angles,    $\hn^\alpha=\hvn \cdot \hvx^\alpha$, between the radial direction and the Cartesian coordinates,  satisfy the orthogonality condition, 
\begin{equation}
\label{eq:northo}
\int \dd \Omega \hn^\alpha \hn^\beta=\frac{4\pi}{3}  \delta_{\rm K}^{\alpha\beta} \; ,
\end{equation} 
where $ \delta_{\rm K}^{\alpha\beta}$  is the Kronecker delta function which equals to unity for $\alpha=\beta$ and vanishes otherwise. 
Let $\xid_{ij}=\la d_i d_j\ra $ and $\xis_{ij}=\la u_i u_j \ra$ denote, respectively, the {\rm data-data}  and the underlying {\it signal-signal} correlation functions\footnote{We shall use the terms ``correlation functions" and ``covariance matrices"  interchangeably.}. The angle brackets imply ensemble average over all possible realizations of the quantity inside the brackets.
Since the  error, $e_i$,  and the signal, $u_i$,  are  uncorrelated, we obtain
\begin{equation}
\label{eq:datasignal}
\xid_{ij}= \sigma_i^2 \delta_{\rm K}^{ij} + \xis_{ij}\; .
\end{equation}
The specification of the target quantity of interest, i.e. the bulk flow, is done via the
 cross correlation 
 $\xidv$,  
 \begin{equation}
\xidv^{\alpha}_i=\la d_i B^\alpha\ra= \la u_i B^\alpha\ra \; .
\end{equation} 
This is the only quantity which involves the definition of the bulk flow. For the definition (\ref{eq:Bdeftwo}) for example we get
\begin{equation}
\label{eq:PII}
\xidv^{\alpha}_i=3\int \dd^3 r g(r)\xis_i(\vr) \hn^\alpha(\vr) \; ,
\end{equation}
where $\xis_i(\vr)=\la u_i u(\vr) \ra$.  
It is also possible to obtain $\xidv^\alpha_i$ for the first definition which  involves the correlation with transverse component of $\vv$ \citep{Gorski1988}, but we do not present this here.

We shall  use boldface to denote vectors and matrices, e.g. $\vd $ represents the explicit notation 
$d_i$ for all $i$ and $\bm{\xid}^{-1} \vd $ is $\sum_j \xid^{-1}_{ij} d_j$. We will alternate between these two conventions, as demanded by the facilitation of mathematical 
manipulations. 

\subsection{Probabilities}
\label{sec:cond}
The joint probability distribution function (PDF) $P(\vB; \vd) $ is our main tool. Specifically, the conditional PDFs $P(\vB|\vd)$  and
$P(\vd|\vB) $ serve as the basis of the methods described here.  
We give a brief summary of the properties of conditional normal  PDFs \citep[e.g.][]{Bertschinger1987,hr91}.
Let $ \vX$  be an $n$-dimensional gaussian random variable divided into two parts $\vX=(\vX_1,\vX_2) $.
The  matrix $\vSigma=\la \vX \vX^{\rm T}\ra$  and its inverse, $\vSigma^{-1}$ are decomposed as
\begin{equation}
\mb{\Sigma}=
\begin{bmatrix}  
\vSigma_{11} & \vSigma_{12} \\
\vSigma_{21} & \vSigma_{22}
\end{bmatrix}\; , \quad 
\mb{\Sigma}^{-1}=
\begin{bmatrix}  
\vSigma^{11} & \vSigma^{12} \\
\vSigma^{21} & \vSigma^{22}
\end{bmatrix}
\end{equation}
where $\vSigma_{kl} =\la \vX_k \vX_l^{\rm T}\ra$
%The inverse of $\vSigma$ is 
%\begin{equation}
%\mb{\Sigma}^{-1}=
%\begin{bmatrix}  
%\vSigma^{11} & \vSigma^{12} \\
%\vSigma^{21} & \vSigma^{22}
%\end{bmatrix}
%\end{equation}
and %%% Inverse Sigma
\begin{eqnarray}
\label{eq:invSigma}
\vSigma^{11}&=&\left(\vSigma_{11}-\vSigma_{12}\vSigma^{-1}_{22}\vSigma^{\rm T}_{12} \right)^{-1}\\
\nonumber &=&\vSigma^{-1}_{11}+\vSigma^{-1}_{11}\vSigma_{12} 
\left(\vSigma_{22}- \vSigma_{12}^{\rm T}\vSigma_{11}^{-1}\vSigma_{12}\right)^{-1}\vSigma_{12} ^{\rm T}\vSigma^{-1}_{11}\\
\nonumber \vSigma^{22}&=&\left(\vSigma_{22}-\vSigma_{12}^{\rm T}\vSigma^{-1}_{11}\vSigma_{12} \right)^{-1}\\
\nonumber &=&\vSigma^{-1}_{22}+\vSigma^{-1}_{22}\vSigma_{12}^{\rm T} 
\left(\vSigma_{11}- \vSigma_{12}\vSigma_{22}^{-1}\vSigma_{12}^{\rm T}\right)^{-1}\vSigma_{12} ^{\rm T}\vSigma^{-1}_{22}\\
\nonumber 
\vSigma^{12}&=&-\vSigma^{-1}_{11}\vSigma_{12}\left( \vSigma_{22}-\vSigma_{12}^{\rm T}\vSigma^{-1}_{11}\vSigma_{12}\right)^{-1}\\
\nonumber&=&  \left(\vSigma^{21}\right)^{\rm T}
\end{eqnarray}
The joint PDF $P(\vX)=P(\vX_1,\vX_1)\propto \exp(-Q/2)$
with
\begin{equation}
\label{eq:Qjoint}
Q=\vX_1^{\rm T} \vSigma_{11}^{-1} \vX_1+\left(\vX_2-\mb{\mu}\right)^{\rm T}\vxi^{-1}\left(\vX_2-\mb{\mu}\right)
\end{equation}
where 
\begin{equation}
\label{eq:mujoint}
\mb{\mu}=\vSigma^{\rm T}_{12} \vSigma^{-1}_{11}\vX_1\quad {\rm and} \quad \vxi=\vSigma_{22}-\vSigma_{12}^{\rm T}\vSigma^{-1}_{11}\vSigma_{12}\; .
\end{equation}
%\begin{eqnarray}
%\label{eq:mujoint}
%\mb{\mu}&=&\vSigma^{\rm T}_{12} \vSigma^{-1}_{11}\vX_1\\
%\nonumber \vxi&=&\vSigma_{22}-\vSigma_{12}^{\rm T}\vSigma^{-1}_{11}\vSigma_{12} \; .
%\end{eqnarray}
From this form of $Q$, the conditional PDF for $\vX_2$ given $\vX_1$ is easily found using Bayes theorem,
$P(\vX_2|\vX_1)=P(\vX_1,\vX_2)/P(\vX_1)\propto \exp(-\Lambda/2)$ where 
\begin{equation}
\label{eq:Lambdacon}
\Lambda=\left(\vX_2-\mb{\mu}\right)^{\rm T}\vxi^{-1}\left(\vX_2-\mb{\mu}\right) \; .
\end{equation} 
\section{Inference based on $P(\vd|\vB)$: generalized ML estimation}
\label{sec:freq}
 An estimate for $\vB$  is obtained by means of Maximum Likelihood (ML) estimation, i.e. by maximizing $P(\vd |\vB)$,  the likelihood for observing the data given  the model.  This is equivalent to finding $\vB$ which renders a minimum in   $\Lambda$ by solving 
$\partial \Lambda/ \partial B^\alpha=0$.
We take $\vX_1=\vB$ to represent the three component of the bulk flow and $\vX_2=\vd$ the $N$-dimensional data vector.  
The specification of the precise definition of the  bulk flow is fixed via the covariance matrix $\vSigma_{12}=\vSigma_{Bd}=\mb{\xidv}=\la\mb{u} \vB\ra $. The matrix  
 $\vSigma_{11}=\vSigma_{BB}$ is a $3\times 3$ matrix corresponding to $\la B^\alpha B^\beta\ra$.  Isotropy implies $\vSigma_{BB}^{\alpha\beta}=\sigma_B^2\delta_{\rm K}^{\alpha\beta}$.  Further, 
$\vSigma_{22}=\vSigma_{dd}= \mb{\xid}$.
%%%%%%% muxi
Therefore, according to (\ref{eq:mujoint}), 
\begin{eqnarray}
\label{eq:muxi}
\mu_i&=&\sum_{\alpha} \sigma_B^{-2} \xidv^\alpha_i B^\alpha\\
\nonumber \xi_{ij}&=&D_{ij}-\sum_{\alpha} \sigma_B^{-2}\xidv^\alpha_i \xidv^\alpha_j \; ,
\end{eqnarray}
 and the minimization of the corresponding expression for $\Lambda$ in (\ref{eq:Lambdacon})   yields 
\begin{equation}
\label{eq:BMLE}
B^{\alpha}=B^\alpha_{_{d|B}}=\sum_\beta ({\cal A}^{-1})^{\alpha\beta} \sum_{i,j}\sigma_B^{-2}\xi^{-1}_{ij} d_j \xidv_i^\alpha 
\end{equation}
where 
\begin{equation}
\label{eq:AMLE}
{\cal A}^{\alpha\beta}=\sum_{i,j}(\sigma_B^{-2})^2\xi_{ij}^{-1}\xidv_i^\alpha \xidv_j^\beta\; .
\end{equation}

\subsection{A Special Case: The standard ML estimation }
As a special case we consider the  velocity model  $\vv=\vB_0=const$, implying
$ u_i=\sum_\alpha \hn^\alpha_i \cdot  B_0^\alpha\;  .
$
 For either    definition of $\vB$ (\ref{eq:Bdefone} $\&$  \ref{eq:Bdeftwo})  and for any $g(r)$, this model gives
 \begin{eqnarray}
 \label{eq:trivialu}
{\cal S}_{ij} &=&\la u_i u_j\ra=\sigma_B^2\sum_\alpha \hn_i^\alpha \hn_j^\alpha\\
\xidv_i^\alpha&=& \la u_i B^\alpha\ra=\hn_i^\alpha \sigma_B^2\\
\mu_i&=& \sum_\alpha \hn_i^\alpha B^\alpha\\
 \xi_{ij}&=&\sigma_i^2 \delta^{\rm K}_{ij} \; ,
 \end{eqnarray}  
 where the last equality is obtained by substituting (\ref{eq:datasignal}) in the expression   (\ref{eq:muxi}) for $\xid_{ij}$.
 Therefore,   (\ref{eq:Lambdacon}) reduces to 
 \begin{equation}
\label{eq:LambdaMLE}
\Lambda=\sum_i\frac{(d_i-\sum_\alpha \hn^\alpha_i B^\alpha)^2}{\sigma_i^2}\; .
\end{equation}
In this expression   the residual between the data and the model for $u_i$
is uncorrelated and is entirely due to observational error and possible very small scale velocity dispersion. 
The solution for $\partial \Lambda /\partial B^\alpha=0$ is %  is $\sum_i w^\alpha d_i$
\begin{equation}
B^\alpha=\sum_\beta({\cal A}^{-1})^{\alpha\beta}\sum_{i=1}^N\ \frac{\hn_i^\beta d_i}{\sigma_i^2}
\end{equation}
where
%%%% Alphabet
 \begin{equation}
 \label{eq:A}
{\cal A}^{\alpha\beta}=\sum_{i=1}^N \frac{\hn_i^\alpha \hn_i^\beta}{\sigma_i^2}\; .
\end{equation}
The summation is over all particles independent of the form of $g(r)$. 
Thus, this special velocity model reproduces the standard ML result  \citep{Kaiser88}. 
  In order to derive estimate for the $\vB$ within a  sphere of radius $R_0$, the actual model has to be modified into 
$\vv(\vr)=\vB_0$ for $r\le R_0$ and $\vv=0$ otherwise.

%
%\section{MLE}
%The standard MLE minimizes 
% The generalization of \ref{eq:LambdaMLE} is best done by considered the conditional probability for $d_i$ 
% given a value for the bulk flow defined in anyway one desires.
%  \begin{equation}
% \label{eq:LambdaCON}
% \Lambda =(\vd - \mb{u}^{\rm m} ) ^{\rm T}{{ \mb{\xi}}}^{-1}(\vd - {\mb{u}}^{\rm m} )
% \end{equation} 
% where 
% \begin{equation}
% { u}^{\rm m}_i=\sum_\alpha \sigma_B^{-2} {\la u_i B^\alpha\ra}  B^\alpha
% \end{equation}
% where $\la B^\alpha B^\beta\ra =\sigma_B^2 \delta_{\rm K}^{\alpha \beta}$ and
% \begin{equation}
% \xi_{ij}=\sigma_i^2\delta_{\rm K}^{ij} + \la u_i u_j\ra - \sigma_B^{-2} \sum_\alpha \la u_i B^\alpha\ra  \la u_j B^\alpha\ra \; .
% \end{equation}
% The angle brackets represent ensemble average.
% 
%
 
\section{Inference based on $P(\vB|\vd)$: Wiener Filtering }
\label{sec:bay}
%%%WF BASICS
We now consider the conditional PDF $P(\vB|\vd)$.
Therefore, we take $\vX_1=\vd$, $\vX_2 =\vB$, $\vSigma_{11}=\mb{\xid}$ and, as before, $\vSigma_{12}=\mb{\xidv}$.  
The corresponding expression  for $\Lambda$ is
$\Lambda=(\vB-\mb{\mu})^{\rm T} \vxi^{-1} (\vB-\mb{\mu}) $ and its minimum is rendered  at 
\begin{equation}
\label{eq:BWF}
B^\alpha=B_{_{\rm B|d}}^\alpha=\mu^\alpha= \sum_{i,j}\xid^{-1}_{ij}  P_j^\alpha d_i\; .
\end{equation}
where the last equality is obtained 
by substituting the relevant quantities in (\ref{eq:muxi}). 
The covariance matrix  $\vxi $ does not appear in  $\vB_{_{\rm B|d}}$, but we give it here for later use
\begin{equation}
\label{eq:xiWF}
\xi^{\alpha\beta}=\sigma_B^{2}\delta_{\rm K}^{\alpha\beta}-\sum_{i,j}\sigma_B^{-2}\xid_{ij}^{-1}\xidv_i^\alpha\xidv_j^\beta\; .
\end{equation}

\subsection{ The relation between $\vB_{_{B|d}} $ and $\vB_{_{d|B}} $ }
Since $P(\vB|\vd)P(\vd)=P(\vd |\vB)P(\vB)$, the expressions (\ref{eq:BWF})  and (\ref{eq:BMLE}) must coincide in the limit of $P(\vB)=const$, i.e. 
for $\sigma_B\rightarrow \infty$. 
%\begin{equation}
%\vB_{_{B|d}}=\vSigma_{Bd}^{\rm T} \vSigma_{dd}^{-1}\vd
%\end{equation}
%%%
To see how this happens we resort to the matrix notation 
and  write ${\cal A}^{\alpha\beta}$ in (\ref{eq:AMLE}) as 
\begin{eqnarray}% 1=B 2=d
\mb{\cal A}&=&\vSigma^{-1}_{BB}\vSigma_{Bd} 
\left(\vSigma_{dd}- \vSigma_{Bd}^{\rm T}\vSigma_{BB}^{-1}\vSigma_{Bd}\right)^{-1}\vSigma_{Bd} ^{\rm T}\vSigma^{-1}_{BB}\\
\nonumber &=&\vSigma^{BB}-\vSigma^{-1}_{BB}\; ,
\end{eqnarray}
where  the equalities are derived using (\ref{eq:invSigma}) with $1\rightarrow B$ and $2\rightarrow d$.
Further the sum over $i,j$ in the solution (\ref{eq:BMLE})  transforms into  
\begin{eqnarray}
&&\vSigma_{BB}^{-1}\vSigma_{Bd}\left(\vSigma_{dd}-\vSigma_{Bd}^{\rm T}\vSigma_{BB}^{-1}\vSigma_{Bd}\right)^{-1}\vd \\
\nonumber &=&\vSigma_{dd}^{-1}\vSigma_{Bd}\left(\vSigma_{BB}-\vSigma_{Bd}^{\rm T}\vSigma_{dd}^{-1}\vSigma_{Bd}\right)^{-1}\vd \\
\nonumber &=& \vSigma_{dd}^{-1}\vSigma_{Bd}\vSigma^{BB}\vd \\
\nonumber &=& \vSigma^{BB} \vB_{_{B|d}} 
\end{eqnarray}
Hence, in matrix notation
(\ref{eq:BMLE})  is 
\begin{equation}
\mb{\cal A}\;  \vB_{_{d|B}}= \vSigma^{BB} \vB_{_{B|d}} 
\end{equation}
With a little more matrix manipulation  we obtain
\begin{equation}
 \vB_{_{d|B}}=\left[\vSigma_{BB}-\left(\vSigma^{BB}\right)^{-1}\right]^{-1}\vSigma_{BB} \vB_{_{B|d}}\; ,
\end{equation}
where the last equality is obtained from the first two lines in (\ref{eq:invSigma}).
The contribution of $\left(\vSigma^{BB}\right)^{-1}$ is 
negligible compared to $\vSigma_{BB}$  in the limit of $\sigma_B\rightarrow \infty$. This is demonstrated  by proving, 
 \begin{equation}
 \la \vB^{\rm T}\left(\vSigma^{BB}\right)^{-1}\vB\ra\ll \la \vB^{\rm T} \vSigma_{BB} \vB\ra =3\sigma_B^4\; . 
 \end{equation}
We write  $d_i=\mu_i+\Delta_i$ where $\mu_i$ is the mean value of $d_i$ given $\vB$ (cf. eq.~\ref{eq:muxi}) and $\Delta_i $ is a random variable uncorrelated with 
$B^\alpha$. Hence, in the limit of very large $\sigma_B$, 
 \begin{equation}
 \label{eq:Dapprox}
 \xid_{ij}=\la d_i d_j \ra \approx \sum_\alpha \sigma_B^{-2}  \xidv_i^\alpha \xidv_j^\alpha \; .
 \end{equation}
Substituting  $\left(\vSigma^{BB}\right)^{-1}=\vSigma_{BB}-\vSigma_{Bd}^{\rm T}\vSigma_{dd}^{-1}\vSigma_{Bd}$ 
 corresponding to $\sigma_B^2\delta_{\rm K}^{\alpha\beta}-\sum_{i,j}\xid^{-1}_{ij} \xidv_i^\alpha \xidv_j^\beta$, 
 the l.h.s of the inequality becomes
 $3 \sigma_B^2-\sum_{\alpha,\beta}\sum_{i,j}\xid^{-1}_{ij} \xidv_i^\alpha \xidv_j^\beta \la B^\alpha B^\beta\ra\approx 0$, for 
   the approximate $\xid$ given in (\ref{eq:Dapprox}). 

%% 3D
\subsection{Equivalence with $\vB$ from a reconstruction of the full 3D peculiar velocity field}
 \label{sec:fullv}
 
 \cite{Hoffman15} use the  WF methodology \citep{Zh95}  to derive the full 3D peculiar velocity field in space, $\vv(\vr)$, from the measured 
 radial peculiar velocities in the CF2 catalog \citep{CF2}. They then compute the 
 bulk flow as defined in (\ref{eq:Bdefone}) for a Top-Hat window, i.e. the mean $\vv$ within a sphere centered on the observer. They could have also 
 computed the bulk flow according to definition  (\ref{eq:Bdefone}).
 We will  show now that $\vB$ derived from this procedure is (unsurprisingly)  entirely equivalent to    $\vB_{_{\rm B|d}}$.
 The full field is obtained as the one which maximizes the probability $P(\vv(\vr)|\vd)$. Substituting  $\vX_1=d$ and $\vX_2=\vv$ in  (\ref{eq:mujoint})
 and minimizing   $\Lambda=-2\ln P +const$ in (\ref{eq:Lambdacon}), we obtain
 \begin{equation}
 \label{eq:fullv}
 v^\alpha(\vr)=\sum_{i,j} \xid^{-1}_{ij} \la u_i v^\alpha(\vr)\ra d_j\; .
 \end{equation} 
 Substituting this expression into either of the bulk flow definitions  (\ref{eq:Bdefone}) \&  (\ref{eq:Bdeftwo}), 
 leads to the same expression as $\vB_{_{B|d}} $ in (\ref{eq:BWF}).
  
\section{Minimum variance estimation}
\label{sec:MV}
So far we have considered estimates which maximize  normal conditional  PDF.  
The minimum variance  (MV) approach described below yields identical results, but is not restricted to normal PDFs \citep[cf.][for a detailed account]{Zh95}. 
A MV  estimate of $\vB$ is sought as a linear combination of the data, 
\begin{equation}
\label{eq:Bmv}
\vB_{_{\rm MV}}= {\bm w} {\bm d} \equiv \sum_{\alpha=1}^3\sum_{i=1}^N \hvx^\alpha w_i^\alpha d_i\; .
\end{equation} 
%%%%%% LAMBDA
The weights $w_i$ are found by minimizing the variance, $\Lambda_{\rm MV}$, of the residual between $\vB_{_{\rm MV}}$ and  all of its possible realizations, 
\begin{eqnarray}
\label{eq:Lambdamv}
\Lambda_{_{\rm MV}}&=&\langle \left(\vB-{\bm w}{\bm d}\right)^2\rangle\\
\nonumber &=& \langle \vB^2\rangle - 2 \mb{\xidv}  \mb{w} + \mb{w}^{\rm T}   \mb{D}\mb{w}\; .
\end{eqnarray}
The  minimum in $\Lambda{_{\rm MV}}$ is obtained  at 
\begin{equation}
 \mb{w}_{_{\rm MV}}= {\bm \xid}^{-1}{\bm \xidv}\; .
\end{equation}
Implementing this result into (\ref{eq:Bmv}) yields  
$
B^\alpha_{_{\rm MV}}=\sum_{i,j}{\xid}^{-1}_{ij} \xidv^\alpha_{j} d_j $. Therefore, referring to (\ref{eq:BWF}), we find $\vB_{_{\rm MV}}=\vB_{_{B|d}}$.

%%%%%%%%%%%%%%
%%%%%%%%%%%%%%

\subsection{Constrained Minimum Variance }
\label{sec:feld}
%Agarwal2012
One may impose   constraints to any of the above procedures,  as done by  
\cite{feldwh10}  for the MV  estimator.
Their constraint is that 
  if the 3D velocity  field is $\vv=\vB_0=const$, then (\ref{eq:Bmv}) must reproduce $\vB_0$ in the absence of observational errors.
Substituting  $d_i=u_i=\sum_\alpha  B_0^\alpha \hn_i^\alpha$ in  (\ref{eq:Bmv}), the constraint implies
\begin{equation}
B_0^\alpha=\sum_{i,\beta} w_i^\alpha   B_0^\beta \hn_i^\beta\; ,
\end{equation}
 must hold for any $\vB_0=const$,  yielding
\begin{equation}
\label{eq:const}
\sum_{i=1}^N w_i^\alpha \hn_i^\beta=\delta_{\rm K}^{\alpha \beta}\; .
\end{equation}
This  constraint  is incorporated   by modifying (\ref{eq:Lambdamv}) into 
\begin{equation}
\label{eq:lambdacon}
\Lambda_{_{\rm CMV}}= \la (B^\alpha-\sum_i w^\alpha_i d_i)^2\ra+\sum_{\beta} \lambda^{\alpha\beta}\left(\sum_i w_i^\alpha \hn_i^\beta-\delta_{\rm K}^{\alpha \beta}\right)\; ,
\end{equation}
to be minimized with respect to the Lagrange multipliers, $\lambda^{\alpha\beta}$ as well as $w_i^\alpha$.
The solution for the minimum point is  \citep[c.f. eqs. 8-10 in ][]{Agarwal2012},
%\begin{equation}
%w^\alpha_i = \sum_j D^{-1}_{ij}\left( \langle d_i B^\alpha \rangle - {1\over 2}\sum_{\beta=1}^3 \lambda^{\alpha\beta}\hn^{\beta}_ j\right) ,
%\label{eq:weight}
%\end{equation}
%and
%\begin{equation}
%\lambda^{\alpha\beta}= \sum_{\gamma=1}^3\left[ (M^{-1})^{\alpha \gamma}\left(\sum_{i,j}D^{-1}_{ij}\langle d_iB^\gamma \rangle 
%\hn^\beta_j- \delta^{\beta \gamma}_K\right)\right],
%\end{equation}
%
\begin{eqnarray}
w^\alpha_i &=& \sum_j D^{-1}_{ij}\left( \xidv_i^\alpha- {1\over 2}\sum_{\beta=1}^3 \lambda^{\alpha\beta}\hn^{\beta}_ j\right) \\
\nonumber \lambda^{\alpha\beta}&= &\sum_{\gamma=1}^3\left[ (M^{-1})^{\alpha \gamma}\left(\sum_{i,j}D^{-1}_{ij}\xidv_i^\gamma
\hn^\beta_j- \delta^{\beta \gamma}_K\right)\right],
\label{eq:weight}
\end{eqnarray}
where
\begin{equation}
M^{\alpha\beta} = {1\over 2}\sum_{i,j} D^{-1}_{ij}\hn^\alpha_i \hn^\beta_j \; .
\end{equation}

%%% w' in SFW
\cite{feldwh10}  adopt  the following scheme to approximate $\xidv_i^\alpha$ (in \S \ref{sec:computing} where we outline an alternative scheme). They write
\begin{equation}
\label{eq:dU}
\langle d_iB^\alpha\rangle = \sum_{j^\prime=1}^{N^\prime} {w^\prime}^\alpha_{j^\prime} \langle {d_i u_{j^\prime}}\rangle
\end{equation}
where
\begin{equation}
{w^\prime}^\alpha_{i^\prime} =  \frac{1}{N^\prime}\sum_{\beta=1}^3 ({A^\prime} ^{-1})^{\alpha \beta}\hn^\beta_{i^\prime}
\end{equation}
are the weights of an  isotropic survey of $N^\prime$ exact radial velocities $u_{i^\prime}$ measured at random positions ${\vr}_{i^\prime}^\prime$ having the radial distribution fixed by $g(r)$, and 
\begin{equation}
{A^\prime}^{\alpha \beta}= \frac{1}{N^\prime} \sum_{i^\prime=1}^{N^\prime} \hn^\alpha_{i^\prime} n^\beta_{ i^\prime}\; .
\end{equation}
In the continuum limit of $N^\prime \rightarrow \infty$, $A^{\prime \alpha\beta}=\int_{4\pi} \dd \Omega r^2 \dd r g(r)\hn^\alpha(\hvn) \hn^{\beta}(\hvn)/\int \dd \Omega r^2 \dd rg(r) =
1/3\delta_{\rm K}^{\alpha\beta}$. Hence, ${w^\prime}^\alpha_{i^\prime}=3\hn_{i^\prime}^\alpha$ and (\ref{eq:dU}) becomes
\begin{equation}
\langle d_iB^\alpha\rangle=3\int\dd \Omega {r^\prime}^2 \dd r^\prime g(r^\prime) \hn^\alpha(\hvn^\prime)\la u_i u(\vr^\prime)\ra\; ,
\end{equation}
which  clearly  coincides with (\ref{eq:PII}).
 %%% NEXT subsection
\subsection{No velocity correlation \& standard ML}
  We discuss now the case where  $D_{ij}=(\sigma_v^2 +\sigma_{0i}^2)\delta^K_{ij}$. 
According to (\ref{eq:dU}), $\la d_i B^\alpha\ra =0$ since  overlap between the data and the ideal sample is zero.
Therefore, 
\begin{eqnarray}
M^{\alpha\beta}&=&\frac{1}{2}\sum_{i}\sigma_i^{-2}\hn^\alpha_i n^\beta_i\\
\nonumber &=& \frac{1}{2} A_{\alpha \beta}\\
\lambda^{\alpha \beta}&=&  -(M^{-1})^{\alpha \beta}\\
w^\alpha_i&=& \sum_{\beta=1}^3 (A^{-1})^{\alpha \beta} \frac{\hn^\beta_i}{\sigma_i^2}
\end{eqnarray}
where 
\begin{equation}
A^{\alpha \beta}=\sum_{i=1}^{N}\frac{\hn^\alpha_i \hn^\beta_ i}{\sigma_i^2} \; .
\end{equation}
Therefore,
\begin{equation}
\mb{w}=\mb{w}_{_{\rm ML}} \; ,
\end{equation}
i.e.  the weights reduce exactly to the ML weights.  In contrast, $\vB_{_{\rm B|d}}$ approaches zero for  uncorrelated velocities, as expected.

\subsection{ Generalization of the constraint}
%% Perhaps move somewhere else
The constraint imposed above yields weights that recover a constant velocity field in the case of perfect data. 
The data points are not uniformly distributed and one may argue that this constraint is actually inappropriate since 
there will always be leakage from other modes of the velocity field. 

If desired, other constraints which may be more realistic could be imposed. 
In particular, the radial velocity field could be decomposed into a functional basis of products of spherical harmonics and spherical Bessel functions \citep{Regos1989,Scharf1993,Fisher95b} .  If the velocity field is fully specified by a dipole term then we could write $u(\vr)=\sum _\alpha a^\alpha \psi(r)\hn(\vr)^\alpha $ where $a^\alpha$ are the expansion coefficients and 
$\psi$ represent  the radial functional basis, e.g.  a Bessel function with wavenumber $k$ (see \S\ref{sec:computing} for details). The constraint is that for the ideal case of this form of the velocity field, the bulk flow is correctly recovered. 
Adopting the second $\vB$ definition in (\ref{eq:Bdeftwo}), the constraint becomes
\begin{eqnarray}
\nonumber 
&& {3}\int \dd^3 r g(r)  \sum _\beta a^\beta \psi(r)\hn^\beta (\vr)\hn^\alpha(\vr)\\
&=&\sum_iw_i^\alpha \sum_\beta a^\beta \psi_i\hn^\beta_i
\end{eqnarray}
 Performing the angular integration and remembering  that the constraint must hold for any $a^\beta$ we get 
 \begin{equation}
\sum_i w_i^\alpha \psi_i \hn_i^\beta =\Psi \delta_{\rm K}^{\alpha\beta} 
 \end{equation}
 where $\Psi=4\pi\int \dd r r^2 g(r) \psi(r) $. Substituting  this equality as our  constraint in $\Lambda_{\rm CMV}$  (instead of
 \ref{eq:const}),   the corresponding weights    are  given by  (\ref{eq:weight}) but with the modification,
 $\hn_i^\alpha\rightarrow  \psi_i \hn_i^\alpha$ and $\delta_{\rm K}^{\alpha\beta}\rightarrow  \delta_{\rm K}^{\alpha\beta}\Psi$

\subsection{Alternative method for imposing constraints}
\label{sec:alt}
Constraints can be imposed by modifying the conditional PDF $P(\vB|\vd)\propto \exp(-\Lambda/2)$. This is done 
adding either a quadratic or linear term in $\vB$ to $\Lambda$.  Here we  present only the quadratic modification since 
it leads to a particularly elegant result. 
We write 
\begin{equation} 
\Lambda_{\rm mod}=\left(\vB-\vB_{_{B|d}}\right)^{\rm T} \vxi^{-1}\left(\vB-\vB_{_{B|d}}\right)  +\vB^{\rm T} (\vxi \vDelta)^{-1} \vB\; . 
\end{equation}
as minus twice the log of the modified PDF, where $\vB_{_{B|d}}$ and $\vxi$ are given in (\ref{eq:BWF}) and (\ref{eq:xiWF}), respectively.
The $3\times3$ matrix, $\vDelta$,  is derived by demanding that $\vB$ which renders a minimum in $\Lambda_{\rm mod}$ also satisfied the desired constraint. 
For the constraint of \cite{feldwh10} of $ \vB=\vB_0$ for $d_i=\hvn\cdot \vB_0$ we find $(\Delta^{-1})^{\alpha \beta}=\delta_{\rm K}^{\alpha \beta}-\sum_{ij} D^{-1}_{ij}\xidv_i^\alpha \hn^\beta$ and the solution $\vB_{_{\rm mod}} $ is given by, 
\begin{equation}
\mb{ W} \vB_{_{\rm mod}} =\vB_{_{B|d}}\; ,
\end{equation}
where ${ W}^{\alpha \beta}=\sum_i w^\alpha_{_{\rm MV}i}\hn_i^\beta$. Since $B^\alpha_{_{B|d}}=\sum_i w^\alpha_{_{\rm MV}i}d_i$, the result clearly 
reproduces $\vB_0$ for $d_i=\sum_\alpha \hn^\alpha_i B_0^\alpha $.

%%%%%%%%%%%%%%%XXXXXXX
% METHODOLOGY
\section{Computing the covariance matrices}
\label{sec:computing}
% define density contrast in notations and preliminaries
In a cosmological model with gaussian initial conditions, all statistical properties of the linear density contrast field, $\delta$ are fully specified by the 
 linear density power spectrum, $P(k)$,
defined via 
$\la \delta_{\vk}\delta^*_{\vk'} \ra= (2\pi)^3 \delta^{^{\rm Dirac}}(\vk-\vk')P(k)$ where
$\delta_{\vk}$ is the Fourier transform of $\delta(\vr)$ and $\vk$ is the wavenumber. 
Assuming potential flow, the velocity statistics  can readily be related to the density field through the  velocity-density relation $\delta=-f(\Omega_{\rm m})\grad \cdot \vv$ \citep{Peeb80}, where $f(\Omega_m)$ is the linear growth factor and $\Omega_{\rm m}$ is the mass density parameter.
All relevant correlation functions  can be expressed in terms of definite integrals involving these power spectra. 
The integrals must be evaluated numerically with lower and upper bounds  on $k$ corresponding to small scales out to horizon scales. The integrals are typically  cumbersome with highly oscillatory functions (spherical Bessel) appearing in their integrands. 
Instead of direct numerical integration, we propose evaluating the integrals using the formalism 
 developed by \cite{Fisher95b} and this section heavily relies on that paper. 
 Radial peculiar velocities are conveniently expressed in terms of spherical Bessel functions, $j_l$, and spherical Harmonics, $Y_{lm}$.
 To do that we imagine a very large spherical volume or radius $R_{\rm max}$ which entirely encompasses the data and is also sufficiently large that 
 the effects boundary conditions are small. 
 Expressing the density contrast as 
 \begin{equation}
 \delta(\vr)=\sum_{n=1}^{n_{\rm max}} \sum_{l=0}^{l_{\rm max}}  \sum_{m=-l}^{+l}C_{ln}\delta_{lmn}j_l(k_n r)Y_{lm}(\hvn) 
 \end{equation}
 where the wavenumbers   $k_n$ are fixed by the  boundary conditions at $R_{\rm max}$ \citep{Fisher95b}. We advocate the 
 boundary conditions which yield gravity potential decaying as $r^{-l}$ for $r>R_{\rm max}$. Since $j_l Y_{lm}$  are eigenfunctions of the 
 Laplacian, these boundary conditions are equivalent to finding $k_n$ which satsify  $\dd \ln j_l(k_n r)/\dd \ln r |_{R_{\rm max}}=-(l+1)$. 
 The numbers  $C_{ln}$ are  fixed entirely   by the boundary conditions and are given in Table A1 of \cite{Fisher95b}.
 The density coefficients are 
 \begin{equation}
 \delta_{lmn}=\int_{r<R_{\rm max}} \dd^3 r \delta(\vr) j_l(k_n r) Y_{lm}(\hvn)\; .
 \end{equation}
 Using the linear velocity-density relation one obtains {\citep{Regos1989,Scharf1993,Fisher95b}, 
\begin{equation}
u(\vr)=f\sum_{l,m,n}C_{ln} \delta_{lmn}\frac{ j'_l(k_n r)}{k_n} Y_{lm}(\hvn)\; .
\end{equation}
 where $j_l'(z)=\dd j_l(z)/\dd z |_{z=k_n r)}$. For velocities measured with respect to the Local Group frame the $l=1$ term should be modified by subtracting $1/3$ from 
 $j'_1(k_n r)$.
  Using 
\begin{equation}
\la \delta_{l_1 m_1 n_1 } \delta^*_{l_2m_2n_2}\ra = P(k_{n_1}) C_{l_1n_1}^{-1} \delta^{K}_{n_1n_2} \delta^{K}_{l_1l_2} \delta^{K}_{m_1m_2}\; ,
\end{equation}
this representation for $u(\vr)$ yields
\begin{eqnarray}
\xis(\vr_i,\vr_j)&&\\
\nonumber &=&f^2\sum_{l,n} \frac{P(k_n)}{k_n^2}  j'_l(k_n r_i) j'_l(k_n r_j)Y_{lm}(\hvn_i)Y^*_{lm} (\hvn_j)\\
\nonumber &=&f^2\sum_{l,n}\frac{2l+1}{4\pi}  \frac{P(k_n)}{k_n^2}  j'_l(k_n r_i) j'_l(k_n r_j)P_l(\hvn_i\cdot \hvn_j)
\end{eqnarray}
where $P_l$ is Legendre polynomial of order $l$ and we have used  $4\pi/(2l+1)\sum_m Y_{lm}(\hvn_i) Y^*_{lm}(\hvn_j)=P_i(\hvn_i\cdot \hvn_j)$.
From $\xis$, the cross correlation can easily be computed $\xidv$. Using the expression (\ref{eq:PII}) which is appropriate for  for definition (\ref{eq:Bdeftwo})
 we get\footnote{The following integrals have been used
$ \int \dd \Omega \hn^z Y_{10}=\frac{1}{2} \sqrt{\frac{3}{\pi}} \frac{4\pi}{3} $,
$\int \dd \Omega \hn^x Y_{11}=\frac{-1}{2} \sqrt{\frac{3}{2\pi}} \frac{4\pi}{3} $,
$\int \dd \Omega \hn^x Y_{1 -1}=\frac{1}{2} \sqrt{\frac{3}{2\pi}} \frac{4\pi}{3} $,
$\int \dd \Omega \hn^y Y_{11}=\frac{-i}{2} \sqrt{\frac{3}{2\pi}} \frac{4\pi}{3} $ and
$\int \dd \Omega \hn^y Y_{1 -1}=\frac{-i}{2} \sqrt{\frac{3}{2\pi}} \frac{4\pi}{3} $}
\begin{eqnarray}
\xidv_i^\alpha&&\\
\nonumber &=&
{3f^2\hn^\alpha_i}\sum_{n} \frac{P(k_n)j'_1(k_n r_i)}{k_n^2}  \int \dd r r^2 g(r) j'_1(k_n r)\; .
\end{eqnarray}
 
\section{Discussion}
\label{sec:disc}
An aim of this contribution is to show that different approaches to the determination of $\vB$ are tightly related. 
All methods yield something which mimics the behavior of the theoretical bulk flow. 
The  non-trivial challenge is to contrast any estimate with  predictions of cosmological  model.
 Systematic  errors related to observables of galaxy properties, contaminate various methods in different 
ways. More distant galaxies are more prone to these systematics, therefore, consistency checks must be performed by applying the methods  subsamples of the data. Furthermore,  spatial coverage of the data depends on  galaxy-type and unavoidable cuts on 
the observables. Thus, methods should be tested on mock galaxy catalogs that mimic the observations in as much as possible, including the observed 
large scale structure. The constrained simulations  \citep{Sorce2013} designed to match the low redshift large scale structure, can potentially  be highly beneficial if  
combined with galaxy formation models. 

We have also emphasized  the differences between the methods. \cite{Hoffman15} provide an estimate for the bulk flow as defined in (\ref{eq:Bdefone}), while 
\cite{feldwh10} aim at $\vB$ given by  (\ref{eq:Bdeftwo}). These two definitions of $\vB$ do not coincide \citep{Nusser2014a}.
and should not produce identical results. Nonetheless, both methods can easily be adapted to any of the  two definitions. 

In \S\ref{sec:freq} we derive a new $\vB$ estimator which naturally arises from the likelihood $P(\vd|\vB)$. This is a generalization of the standard ML of \cite{Kaiser88}, but 
incorporates velocity modes beyond the (constant) bulk flow of the entire  data catalog. Essentially, this estimator is equivalent to marginalizing $P(\vv_{_{\rm model}}|\vd)/P(\vB)$ over 
all modes of a general velocity model, $\vv_{_{\rm model}}$, excluding  a mode corresponding a global bulk flow . Another new estimator is given in \S\ref{sec:alt}. This estimator 
incorporates the \cite{feldwh10} constraint in a new and simple way. But we emphasize that in a Baysian approach, the prior  as expressed in $\xis$ and $\xidv$  correlation functions. 
already contains all missing information. Additional  constraints (not based on extra  information) are either statistically redundant or  incompatible with the model prior.
For example, requiring the estimator to  produce $\vB_0=const$   for  the input data $d_i=\hvn_i \cdot \vB_0$ (no observational errors), ignores 
variations in the velocity field on the scales that are  not probed by the data.

We have refrained from making any quantitative assessment of the difference between methods and 
the bulk flow definitions. 
We hope to do that in the near future as well as comparison applying all methods to observational datat, but  it is not the point of the paper.

\section{Acknowledgments}
This research was supported by the I-CORE Program of the Planning and Budgeting
Committee, THE ISRAEL SCIENCE FOUNDATION (grants No. 1829/12 and No. 203/09), the Asher Space
Research Institute.
\bibliography{/Users/adi/Documents/Bibtex/Velocity.bib}
%\bibliography{/Users/adi/Documents/Bibtex/Radio.bib}
\end{document}